\documentclass[12pt]{article}

\usepackage{amssymb}
\usepackage{color}
\usepackage{graphicx,psfrag}

 \renewcommand{\=}{~=~}

\begin{document}
\begin{center}
   {\Large\bfseries The Ascending Double-Cone}\\
   {\bfseries A Closer Look at a Familiar Demonstration}

\vspace{5mm}

   {\bfseries Sohang C. Gandhi} and {\bfseries  Costas J. Efthimiou}\\
    Department of Physics\\
    University of Central Florida\\
    Orlando, FL 32816
\end{center}

\vspace{5mm}

\begin{abstract}
The double-cone ascending an inclined V-rail is a common exhibit
used for demonstrating concepts related to center-of-mass in
introductory physics courses \cite{PIRAid}. While the conceptual
explanation is well-known---the widening of the ramp allows the
center of mass of the cone to drop, overbalancing the increase in
altitude due to the inclination of the ramp---there remains rich
physical content waiting to be extracted through deeper
exploration. Such an investigations seems to be absent from the
literature.  This article seeks to remedy the omission.
\end{abstract}

\vspace{5mm}

\section{Introduction}
\begin{figure}[h!]
\begin{center}
 \psfrag{f}{$\phi$} \psfrag{T}{\textcolor{red}{$\theta$}}
 \includegraphics[height=4cm]{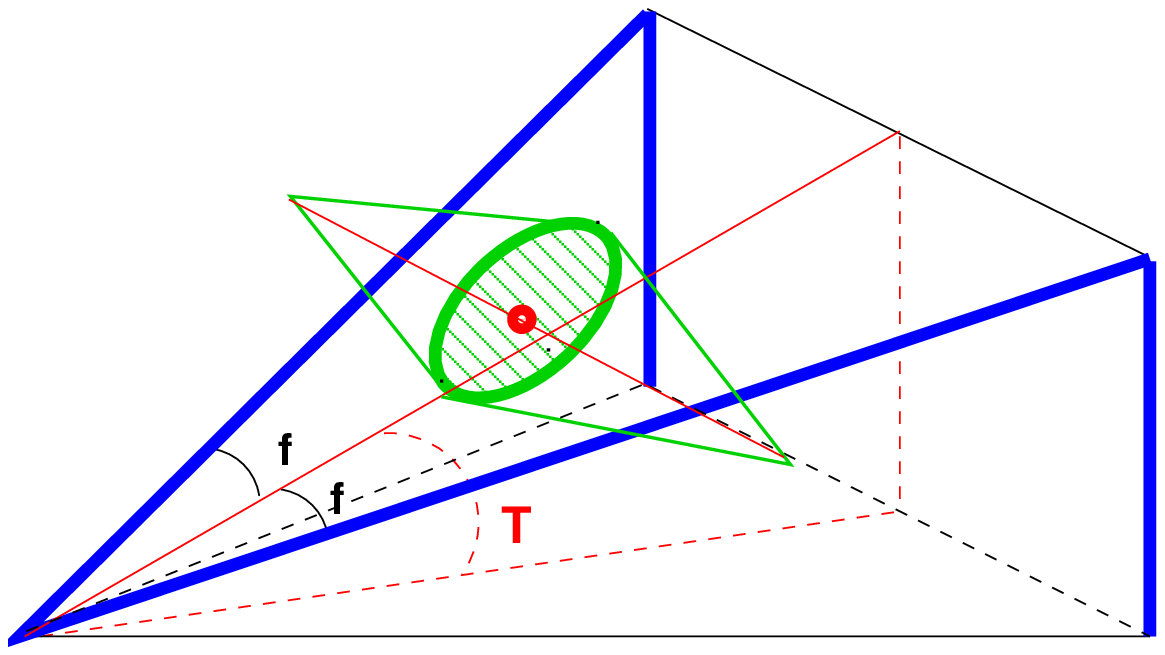}
 \hspace{1cm}
 \psfrag{L}{$L$} \psfrag{s}{\textcolor{blue}{$\psi$}}
 \psfrag{d}{$d$} \psfrag{a}{$a$} \psfrag{L}{$L$}
 \psfrag{CPL}{$CP_{L}$} \psfrag{CPR}{$CP_{R}$}
 \psfrag{P}{$P$}
 \psfrag{CM}{CM}
 \includegraphics[width=8cm]{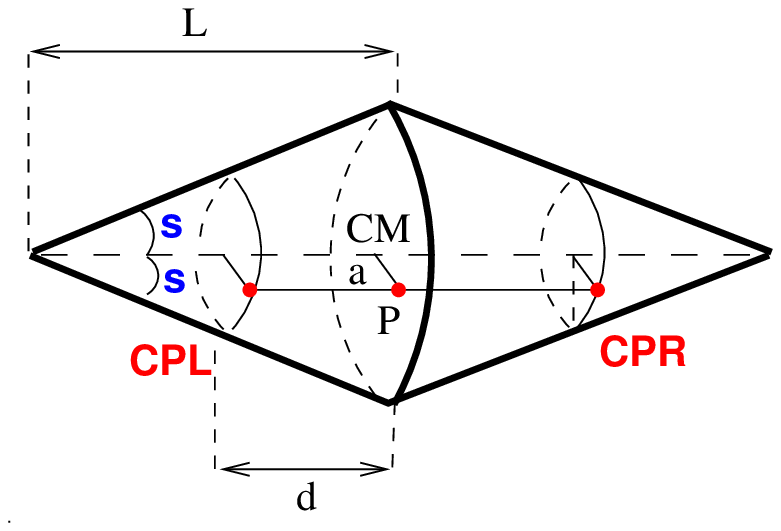}
\end{center}
\caption{{\footnotesize A double cone that rolls up a V-shaped %rail
ramp. The figure shows the principal parameters,
$\theta,\phi,\psi$, determining the geometry of the demo. Also in
the figure, are several quantities used in our study of the
motion. In particular, $a$ is the radius of a cross-section taken
at one of the contact points and perpendicularly to the axis of
the cone.}}
 \label{demo}
\end{figure}

The, familiar, double-cone demonstration is illustrated in figure
\ref{demo} bellow. The set-up consists of a ramp, inclined at an
angle $\theta$, composed of two rails forming a `V' and each
making an angle $\phi$ with their bisector. A double-cone of angle
$2\psi$ is placed upon the ramp. For certain values of the angles
$\theta$, $\phi$, and $\psi$, the double-cone will
spontaneously roll upward.% the ramp.

\begin{figure}[h!]
\begin{center}
\psfrag{CPL1}{\textcolor{red}{$CP_{L1}$}}
\psfrag{CPL2}{\textcolor{red}{$CP_{L2}$}}
\psfrag{CPR1}{\textcolor{red}{$CP_{R1}$}}
\psfrag{CPR2}{\textcolor{red}{$CP_{R2}$}}
\psfrag{a1}{\textcolor{red}{$a_{1}$}}
\psfrag{a2}{\textcolor{red}{$a_{2}$}}
\psfrag{da}{\textcolor{red}{$\Delta a$}}
  \includegraphics[width=9cm]{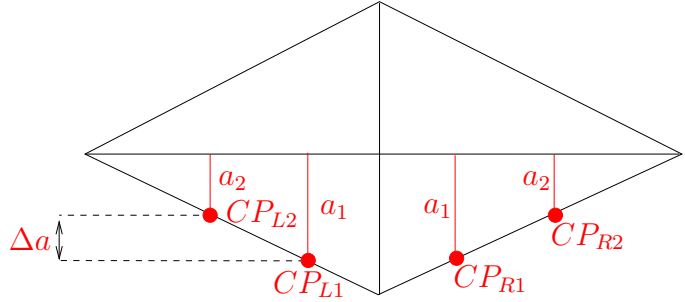}
\caption{{\footnotesize The shifting of the contact
         points as the double-cone rolls uphill. The reader is cautioned. As
         shall be explained, the contact points
         do not lie on the vertical plane that contains the symmetry
         axis of the cone; instead with the symmetry axis of the cone, they define a
         plane that intersects the vertical plane at an angle. See also figure
         \ref{demo}.}}
  \label{dela}
\end{center}
\end{figure}

A qualitative description of the mechanism responsible for the
phenomenon is easily given. As the cone moves up the ramp and
gains altitude $\Delta h$, the distance between the rails
increases and the contact points move nearer the symmetry axis,
thus reducing the elevation of the center-of-mass (CM) relative to
the ramp by an amount $\Delta H$, which is determined by (but not
necessarily equal to) the change of radius $\Delta a$ resulting
from the outward shifting of the contact points (figure
\ref{dela}). The net change of the CM height is thus $\Delta h -
\Delta H$, which may be positive, negative, or zero. When it is
negative the cone rolls uphill since the effect of the widening of
the ramp \textit{overbalances} the increase in altitude.

Despite the above well-known and well-understood explanation for the
motion of the double cone, it seems---to the best of our
knowledge---that no detailed, quantitative account has been written.
Our work seeks to remedy this omission.

\begin{figure}[h!]
\begin{center}
 \psfrag{O}{CM} \psfrag{CP}{$P$}
 \psfrag{t}{\textcolor{red}{\footnotesize $\theta$}}
 \psfrag{a}{\textcolor{red}{$\alpha$}}
 \psfrag{m}{\textcolor{red}{\footnotesize $\alpha-\theta$}}
 \psfrag{N}{\textcolor{red}{$\hat{K}$}}
 \psfrag{A}{\textcolor{red}{$\vec{a}$}}
 \psfrag{B}{}
 \psfrag{s}{$\theta$}
  \includegraphics[width=7cm]{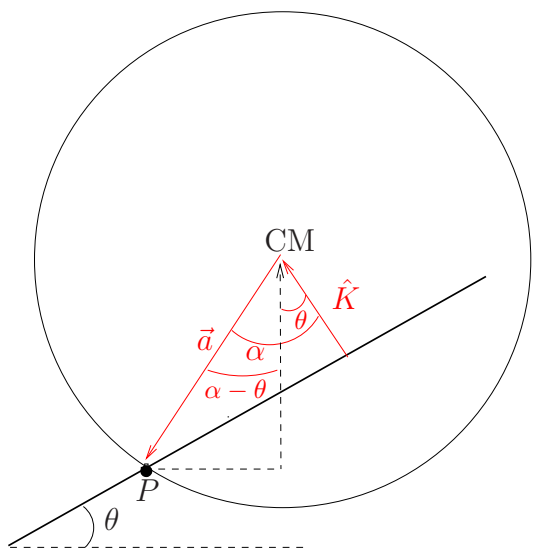}
\caption{ {\footnotesize The circle represents part of the middle
          cross-section of the cone. (See right part of figure
          \ref{demo}.) The solid thick straight  line is the intersection
         of the plane defined by the two rails and a vertical plane perpendicular
         to the axis of the cone.}}
  \label{fig:3}
\end{center}
\end{figure}

Towards this goal we present some notation which will facilitate
the discussions in the following sections. As illustrated in the
right side of figure \ref{demo} and figure \ref{fig:3},  we shall
denote the left and right points of contact as $CP_{L}$ and
$CP_{R}$ respectively, the half length of the cone as $L$, the
distance between the middle cross-section of the cone and the
contact points as $d$, and the distance from the axis of symmetry
to the points of contact as $a$. Also, let $P$ be the point where
the line defined by the two contact points intersects the middle
cross-section of the cone and $\vec a$ be the vector that points
from the center of mass (CM) of the cone to $P$. The magnitude $a$
of $\vec a$ is equal to the radius of the cross-section of the
cone taken at the contact points perpendicularly to the axis of
the cone. When the cone is at the vertex of the ramp, the two
contact points coincide and $a$ has a value of $L\tan\psi$. Also,
we shall denote by $\pi-\alpha$ the angle the vector $\vec a$
makes with the unit normal vector $\hat K$ to the ramp. All this
is better illustrated in figure \ref{fig:3}.

%%%%%%%%%%%%%%%%%%%%%%%%%%%%%%%%%%%%%%%%%%%%%%%%%%%%%%%%%%%%%%%%%%%%%%%%%%
\section{Net Change of Height}

\begin{figure}[htb!]
\begin{center}
\psfrag{X}{$X$} \psfrag{Y}{$Y$} \psfrag{Z}{$Z$}
\psfrag{x}{\textcolor{red}{$x$}} \psfrag{y}{\textcolor{red}{$y$}}
\psfrag{z}{\textcolor{red}{$z$}}
  \includegraphics[width=7cm]{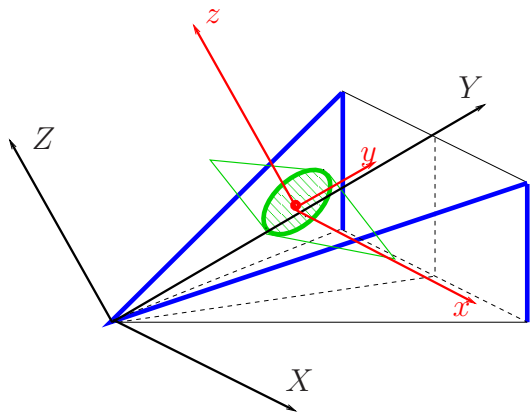}
\end{center}
  \caption{\footnotesize The ramp and CM systems.}
  \label{xyz}
\end{figure}

To carry out our calculation we shall introduce two coordinate
systems: First, a fixed system, which we shall call \textit{the ramp
coordinate system} $XYZ$, with origin at the vertex of the ramp, the
$Y$-axis along the bisector of the rails, the $Z$-axis perpendicular
to the plane defined by the rails and the $X$-axis parallel to the
symmetry axis of the double-cone. We also adopt a CM-system with
axes parallel to those of the ramp system but whose origin is
attached to the CM of the double cone. The systems are shown in
figure \ref{xyz}.

Our experience with uniform spheres, discs, or cylinders rolling
on an incline suggests that a line drawn at a contact point
perpendicularly to the incline on which the objects roll passes
through the CM (or the symmetry axis). Indeed, for these
particular shapes, this is implied by the fact that the incline
must be tangent to the surface of the object at the contact point
and that the radii of a circle are normal to the circle. However,
the conical shape on the V-shaped ramp does not posses the same
geometric property (although a double-cone rolling on two parallel
rails would). Consequently, the contact points, $CP_R$ and $CP_L$,
have a different spatial relation to the CM of the cone (figure
\ref{fig:3}).

Instead of working with the two points $CP_R$, $CP_L$, it is more
convenient to work with the point $P$ defined previously. Notice
that all three points $CP_R$, $CP_L$, $P$ have the same $z$ and
$y$ coordinates and only differ in their $x$ coordinates.  In
fact, due to symmetry---we assume the cone to be placed
symmetrically between the rails so that there is no motion in the
$X$-direction---we are really dealing with a 2-dimensional
problem.

To determine the angle $\alpha$ of figure \ref{fig:3}, we must
find the $y$ coordinate $y_{P}$ in the CM-frame. This will give us
the location of $P$ relative to the CM of the cone. To this end,
we write the equation
$$
   z^2+y^2 -\tan^2\psi\, (L-x)^2 \= 0
$$
that describes the cone and, from this, we construct the surface
\begin{displaymath}
     z \= -\sqrt{(L-x)^2 \, \tan^2 \psi -y^2} ~\equiv~ f(x,y)
\end{displaymath}
which represents the lower right quarter of the cone in the
CM-frame. We then calculate the gradient of the function
$f(x,y)-z$,
\begin{displaymath}
  \nabla z \= {1 \over \sqrt{(L-x)^2 \, \tan^2 \psi -y^2}} \,
  \Big( ~ (L - x)\,\tan^2\psi, ~ y, ~-1 ~\Big)~,
\end{displaymath}
that is normal to the cone. From the definition of the two frames
$xyz$ and $XYZ$ we immediately see that $x=X$ and therefore
$x_{CP}=X_{CP}=d$. Thus, at the right-hand contact point
\begin{displaymath}
     \nabla z\Big|_{CP} \=
     {1 \over \sqrt{(L-d)^2 \, \tan^2 \psi-y_{P}^2}} \,
     \Big( ~ (L - d)\,\tan^2\psi, ~y_{P}, ~-1 ~\Big)~.
\end{displaymath}

 We also construct the unit vector $\hat{u}$ whose direction coincides with
the right-hand rail and points away from the vertex of the ramp.
In other words, the vector  $\hat{u}$ lies in the $xy$-plane and
makes an angle $\phi$ with the $y$-axis; it can therefore be
written as
\begin{displaymath}
   \hat{u} \= (\sin \phi , \cos \phi, 0)~.
\end{displaymath}
As the cone rolls on the rails, the rails are always tangent to
the cone's surface and, therefore, perpendicular to the normal
$\nabla z$:
$$
   \nabla z\Big|_{CP} \cdot \hat u \= 0~,
$$
from which we locate the point $P$,
\begin{eqnarray}
   y_{P} &=& -(L-d)\,\tan^2\psi\,\tan \phi ~.
\label{y}
\end{eqnarray}

\begin{figure}[h!]
\begin{center}
\psfrag{CP}{\textcolor{blue}{$CP$}}
\psfrag{YCP}{\textcolor{blue}{$Y_{CP}$}}
\psfrag{a}{\textcolor{red}{$\alpha$}}
\psfrag{A}{\textcolor{red}{$a$}}
\psfrag{CM}{\textcolor{red}{$CM$}}
\psfrag{ZCM}{\textcolor{red}{$Z_{CM}$}}
\psfrag{YCM}{\textcolor{red}{$Y_{CM}$}} \psfrag{Y}{$Y$}
\psfrag{Z}{$Z$}
  \includegraphics[width=8cm]{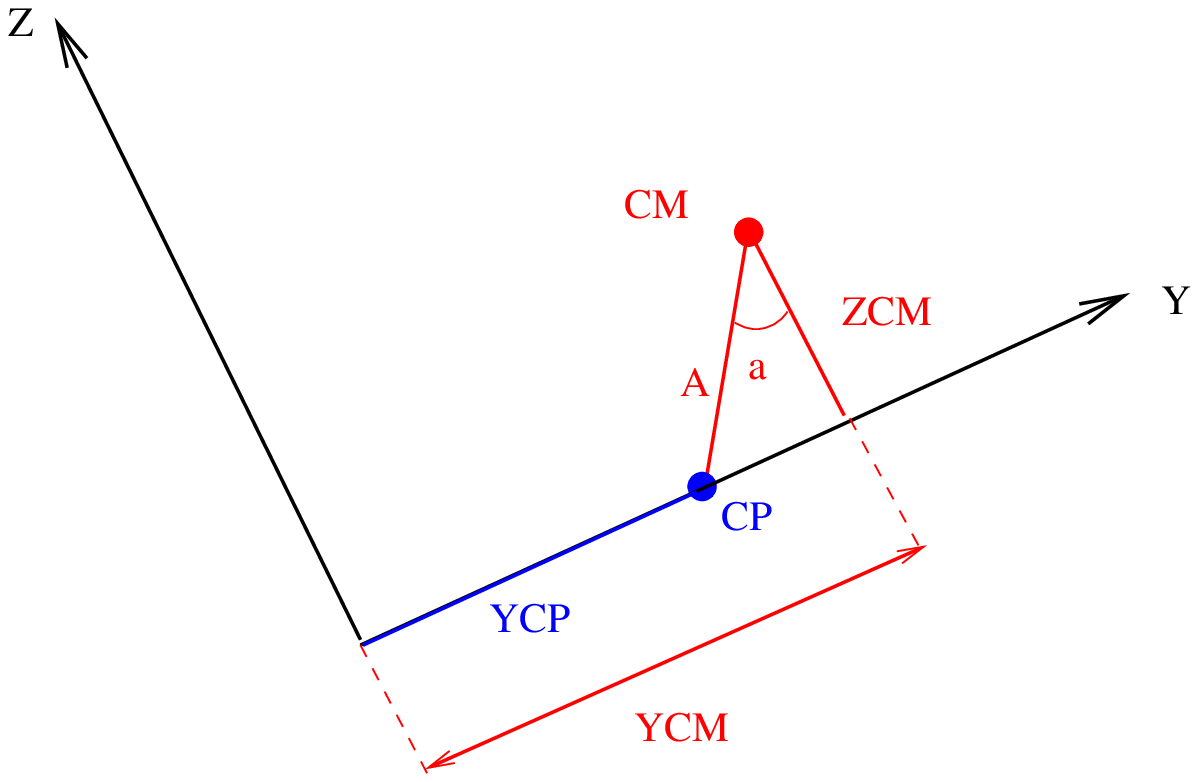}
\hspace{1cm}
\psfrag{f}{$\phi$} \psfrag{YP}{$Y_{P}$} \psfrag{d}{$d$}
\psfrag{P}{$P$} \psfrag{Y}{$Y$}
\includegraphics[width=6.5cm]{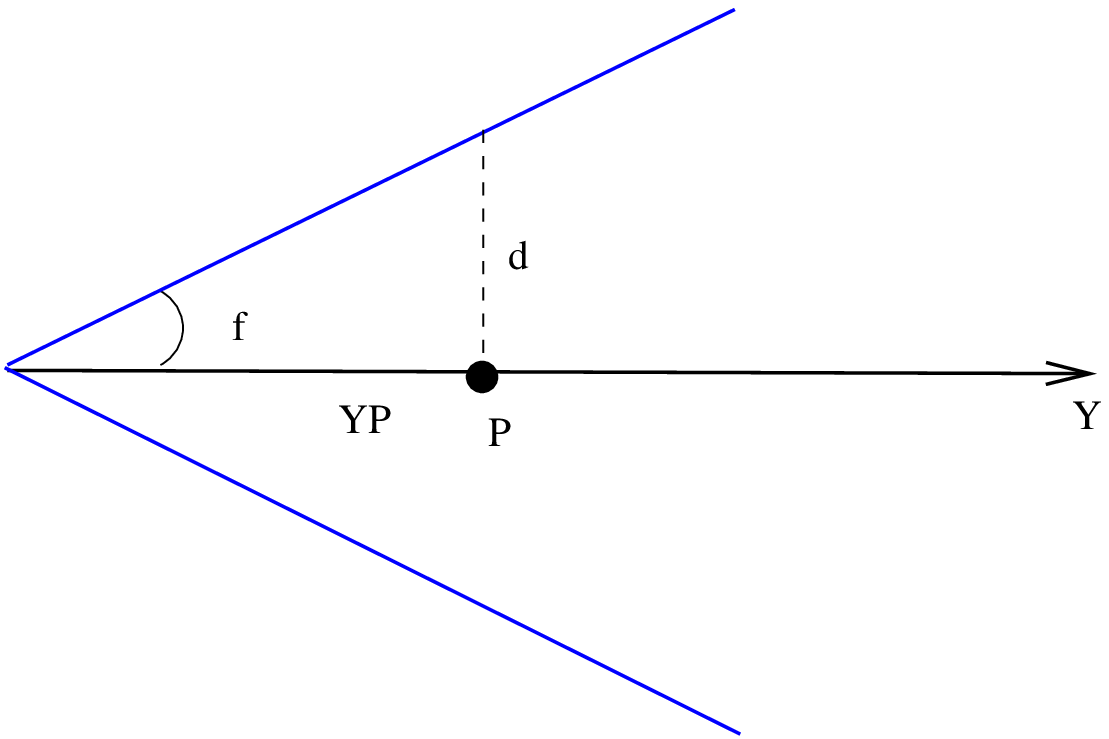}
  \caption{{\footnotesize The relation between some of the parameters
           used in the study of the motion of the cone.}}
  \label{fig:4}
\end{center}
\end{figure}

\begin{figure}[h!]
\begin{center}
 \psfrag{s}{$\psi$}
 \psfrag{s9}{\small ${\pi\over 2} - \psi$}
 \psfrag{p}{$\phi={\pi\over 2}-\psi$}
 \includegraphics{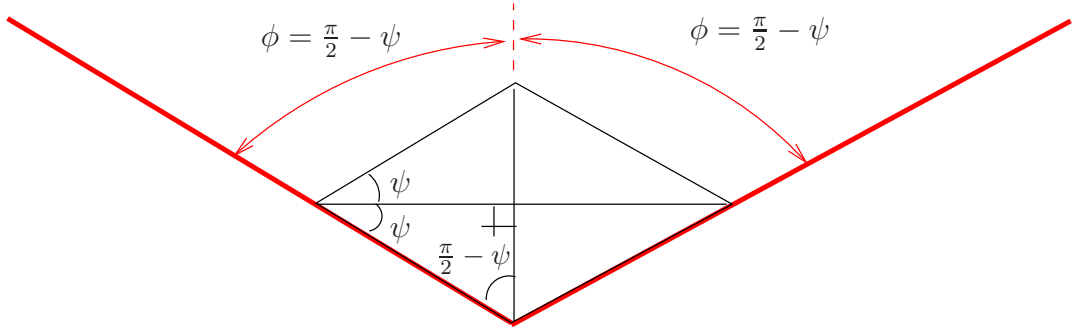}
 \caption{{\footnotesize When a cone of angle $\psi={\pi\over 2}-\phi$, is
           projected on the ramp plane,
          then the cone exactly `fills' the wedge formed by the rails. Increasing the angle
          $\phi$ implies a wedge bigger than what the cone can `fill'.}}
 \label{fallthr}
\end{center}
\end{figure}

From  figure \ref{fig:3}, it can be seen that $y_{P}$ may be
written in terms of $\alpha$ as
\begin{equation}
   y_{P} \= -a\, \sin\alpha ~.
\label{eq:1}
\end{equation}
Also, from figure \ref{fig:4} it can be seen that $a$ can be
written
\begin{equation}
  a \= (L-d)\,\tan\psi~.
\label{afd}
\end{equation}
From equations (\ref{y}), (\ref{eq:1}), (\ref{afd}), we thus find
\begin{equation}
  \sin\alpha \= \tan\phi\,\tan\psi~,
  \label{alph}
\end{equation}
which expresses the angle $\alpha$ in terms of the known angles
$\phi$ and $\psi$. Since $-1\le \sin\alpha\le 1$, there will be a
solution for $\alpha$ if the angles $\phi$ and $\psi$ are such
that $-1\le \tan\phi\,\tan\psi\le 1$. Since  $\phi$, $\psi$  must
all lie in $[0, {\pi\over 2})$, we have that $\tan\phi$,
$\tan\psi$ are non-negative and therefore $0\le
\tan\phi\,\tan\psi$. We use figure \ref{fallthr} to examine the
other side of the inequality. In  the figure, we draw a projection
of the cone onto the ramp plane when $\phi=\pi/2-\psi$. In this
case, the cone exactly `fits' the wedge created by the rails. If
we increase the angle $\phi$, that is if $\phi \geq \pi/ 2-\psi$,
then the cone is allowed to fall through the rails. Therefore, the
demo is well-made if
$$
    \phi ~<~ {\pi\over 2}-\psi~.
$$
For angles in the first quadrant, the tangent function is strictly
increasing. Therefore, the previous condition implies
 $$
   \tan\phi~<~\tan\left({\pi\over 2}-\psi\right)
           \=\cot\psi
           \={1\over\tan\psi}
           ~\Rightarrow~
           \tan\phi\,\tan\psi~<~1~.
 $$

Therefore, for a well-made demo, the inequality
 \begin{equation}
    0~\le~ \tan\phi\,\tan\psi~<~1
 \label{eq:inequality}
 \end{equation}
  is always satisfied and the angle $\alpha$ has a well-defined
value.

We are now able to see the reasoning behind the statement $\Delta
a\ne \Delta H$ made at the introduction. The angle between
$\vec{a}$ and the vertical is $\alpha-\theta$ (figure
\ref{fig:3}).
 Thus, the change in the height $\Delta H$ of the CM relative to the points of contact is
in fact given by  $\Delta a \,\cos(\alpha-\theta)$.  It follows that
the net change in height of the CM is
$$
    \mbox{net~change~of~height} \= \Delta h -\Delta a
    \,\cos(\alpha-\theta)~.
$$

%%%%%%%%%%%%%%%%%%%%%%%%%%%%%%%%%%%%%%%%%%%%%%%%%%%%%%%%%%%%%%%%%%%%%%%%%%%%%%%
\section{A Dynamical Explanation}
With our newfound knowledge regarding the way in which the
double-cone sits on the ramp, we are in a position to provide an
account from a dynamical perspective.

Any solid placed on the ramp will contact it at two points. Since we
assume that the solid rolls without slipping, the contact points
must be instantaneously at rest.  The rigidity of the solid then
implies that all points lying on the line connecting the two contact
points are also at rest instantaneously, and, hence, this line
defines the instantaneous axis of rotation. The torque about this
axis will solely be due to gravity since the frictional forces and
the reaction forces of the ramp are exerted at the axis.

\begin{figure}[h!]
\begin{center}
 \psfrag{CPL}{\textcolor{red}{\scriptsize $CP_L$}}
 \psfrag{CPR}{\textcolor{red}{\scriptsize $CP_R$}}
 \psfrag{t}{\textcolor{black}{\scriptsize $\theta$}}
 \psfrag{A}{\textcolor{black}{\scriptsize $-\vec{a}$}}
 \psfrag{mg}{\textcolor{black}{\footnotesize $m \vec{g}$}}
 \psfrag{T}{$\theta$}
 \psfrag{P}{\scriptsize $P$}
\includegraphics[width=10cm]{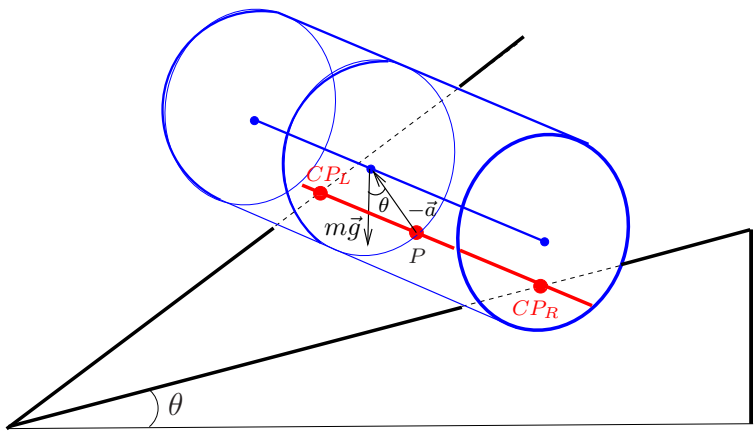}
 \caption{{\footnotesize The resultant torque always pulls a cylinder down the ramp.}}
 \label{cyltor}
\end{center}
\end{figure}

The situation for a cylinder placed on the ramp is shown in figure
\ref{cyltor}. The angle between the weight and the vector $\vec a$,
which enters the torque, is determined by the inclination $\theta$
of the ramp only. Thus the torque resultant of the weight,
$-mga\sin\theta$, always tends the cylinder to roll down the ramp.

\begin{figure}[h!]
\begin{center}
 \psfrag{cpl}{\textcolor{red}{\scriptsize $CP_L$}}
 \psfrag{cpr}{\textcolor{red}{\scriptsize $CP_R$}}
 \psfrag{r}{\textcolor{black}{\footnotesize $-\vec{a}$}}
 \psfrag{mg}{\textcolor{black}{\footnotesize $m \vec{g}$}}
 \psfrag{a}{\tiny $\alpha-\theta$}
 \psfrag{P}{\scriptsize $P$}
\includegraphics[width=10cm]{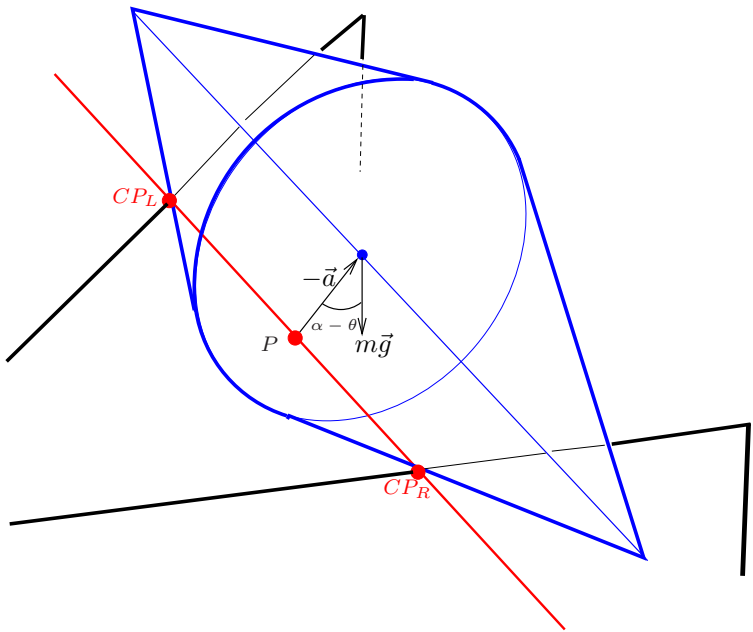}
 \caption{{\footnotesize The resultant torque in the case of a double cone can
           be positive, negative, or zero depending on the value of $\alpha$.}}
 \label{contor}
\end{center}
\end{figure}

The situation for the double-cone is shown in figure \ref{contor}.
The angle between the weight and the vector $\vec a$ which enters
the torque is now $\pi-(\alpha-\theta)$.  The  torque resultant of
the weight, $mga\sin(\alpha-\theta)$,   depends strongly on the
value of $\alpha$, which can change by changing the parameters
$\phi$ and $\psi$.   If $\alpha>\theta$, the resultant torque will
tend the solid up the ramp. We can thus see that it is the unusual
way in which the double-cone sits on the ramp that distinguishes
it from the other solids and provides it the mechanism with which
to roll up the ramp.

%%%%%%%%%%%%%%%%%%%%%%%%%%%%%%%%%%%%%%%%%%%%%%%%%%%%%%%%%%%%%%%%%%%%%%%%%%%%
\section{The Motion of the CM}
Since $Z$ is oriented perpendicular to the plane defined by the
rails and $P$  lies on the ramp plane, $Z_{P}$ is always zero. Thus
$Z_{CM}$ is given by
$$
    Z_{CM} \= a \, \cos\alpha~.
$$
(See figure \ref{fig:4}.)
 If the point $P$ has coordinate $Y_{P}$ in the ramp-frame,
then the half-distance between the rails, $d$, must be given by
\begin{equation}
  d \= Y_{P}\tan\phi~,
\label{dofy}
\end{equation}
as is evident from figure \ref{fig:4}. Using equations
(\ref{dofy}), (\ref{afd}), and (\ref{alph}), we write for $a$,
\begin{equation}
   a \= L\tan\psi-Y_{P}\sin\alpha~.
   \label{aofcp}
\end{equation}
Hence,
\begin{displaymath}
    Z_{CM} \= (L\tan\psi-Y_{P}\sin\alpha) \, \cos\alpha~.
\end{displaymath}
Referring again to figure \ref{fig:4}, we obtain
 $$
  Y_{CM} \= Y_{P} + a\sin\alpha \= Y_{CP} + (L\tan\psi -
  Y_{CP}\sin\alpha)\sin\alpha~,
 $$
 or
 \begin{equation}
 Y_{P} \= {Y_{CM} - L\tan\psi\sin\alpha \over \cos^2\alpha}~.
 \label{cpofcm}
 \end{equation}
Plugging this in to our previous result we obtain
\begin{equation}
  Z_{CM} = \lambda \, Y_{CM} + b~,
  \label{line}
\end{equation}
where $\lambda$ and $b$ are constants determined by the geometry
of the demo:
$$
  \lambda\=-\tan\alpha~,~~~
  b\= L\,\tan\psi(\cos\alpha-\tan\alpha\sin\alpha)~.
$$
Equation (\ref{line}) implies that the CM is constrained to move
along a line on the $zy$-plane with slope $\lambda$ and intercept
$b$.

This result is also made apparent through physical reasoning. The
cone's  motion is that of rigid rotation about the symmetry axis
plus a translational motion of the CM with velocity $\vec v_{CM}$.
Thus, the total velocity of the point  $P$ is $\vec{v}_{rot,P}+\vec
v_{CM}$, where $\vec{v}_{rot,P}$ is its rotational velocity about
the symmetry axis. The rolling without slipping condition implies
that the velocity of the  point $P$ must be zero. Therefore
$$
  \vec v_{CM} \= - \vec{v}_{rot,P}~.
$$
Since $\vec{v}_{rot,P}$ must be perpendicular to $\vec a$, we
conclude that
$$
  \vec{v}_{CM} \perp \vec a~.
$$
Since $\vec{a}$ is fixed in direction and the velocity of the CM
must always lie on the $YZ$-plane (by symmetry), the CM's path is
a line declined (inclined if $\alpha < \theta$) below the
horizontal by an angle $\alpha-\theta$ (see figure \ref{qtounp}).

%%%%%%%%%%%%%%%%%%%%%%%%%%%%%%%%%%%%%%%%%%%%%%%%%%%%%%%%%%%%%%%%%%%%
\section{Energy Analysis}
\begin{figure}
\begin{center}
 \psfrag{Z}{$Z$}\psfrag{Y}{$Y$}
 \psfrag{at}{$\alpha$}
 \psfrag{a}{\textcolor{blue}{$-\vec{a}_0$}}
 \psfrag{q}{\textcolor{red}{$q$}}
\includegraphics[width=7.5cm]{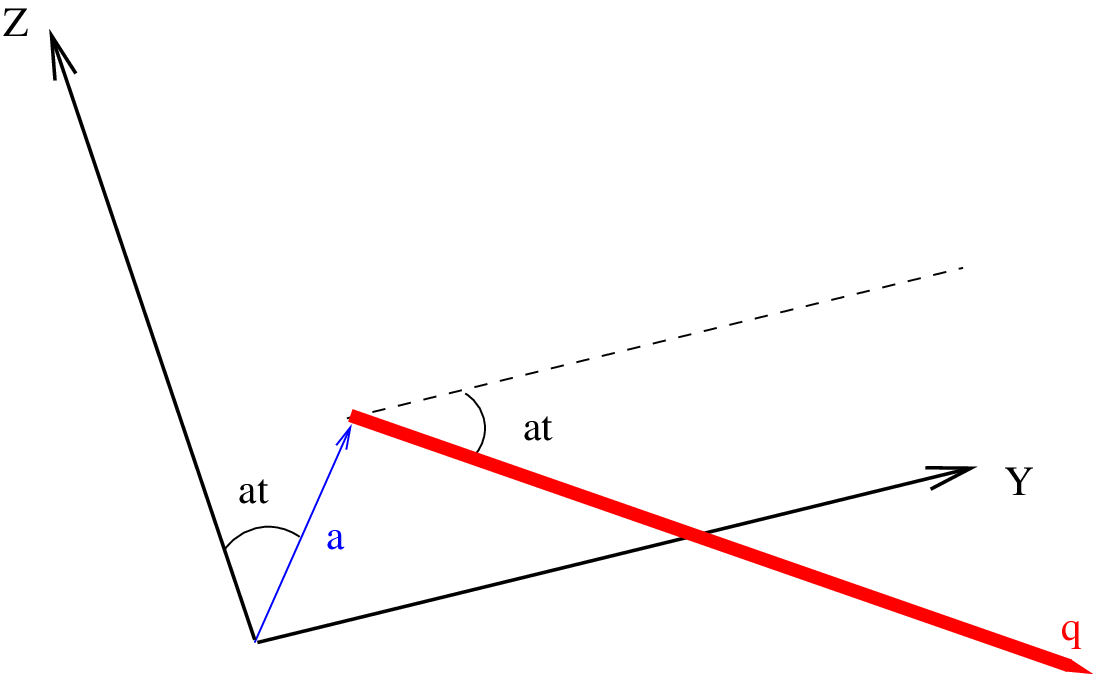}
 \hspace{1cm}
 \psfrag{Z}{$Z$}\psfrag{Y}{$Y$}
 \psfrag{at}{\scriptsize $\alpha-\theta$}
 \psfrag{aa}{\tiny $\alpha-\theta$}
 \psfrag{a}{\textcolor{blue}{$-\vec{a}_0$}}
 \psfrag{q}{\textcolor{red}{$q$}}
 \psfrag{Z'}{$Z'$}
 \psfrag{Y'}{$Y'$}
 \psfrag{t}{$\theta$}
\includegraphics[width=7.5cm]{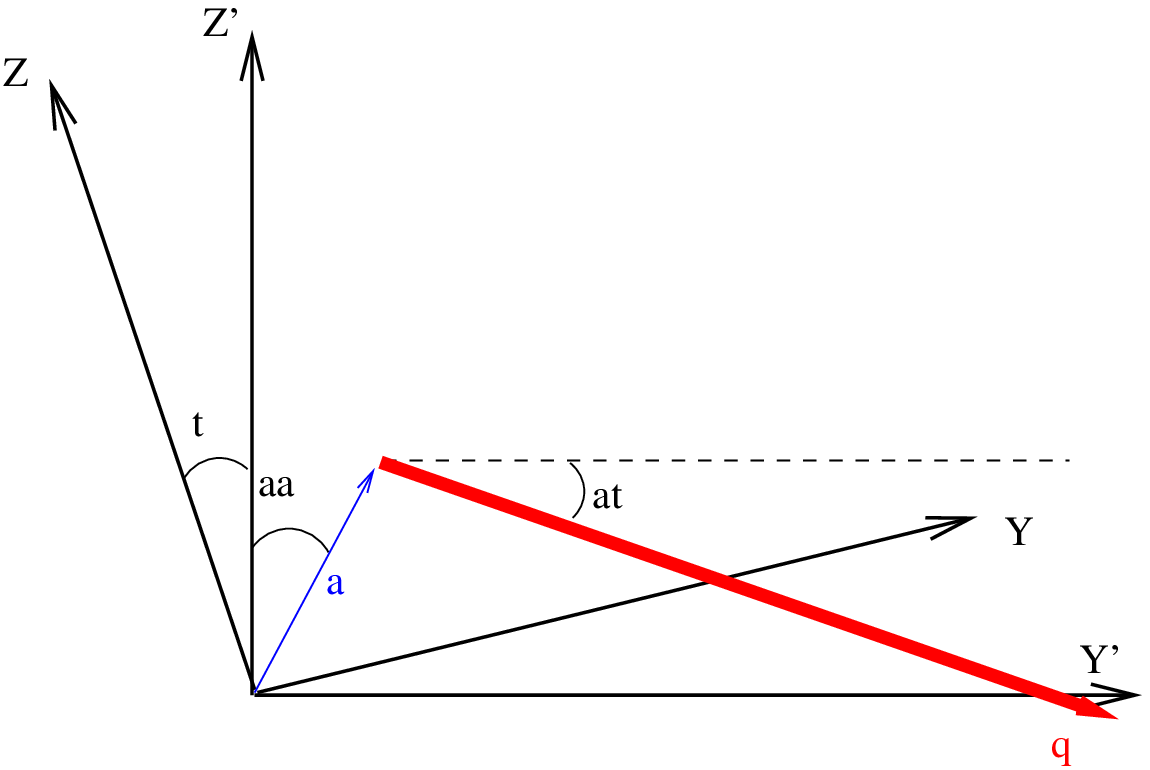}
 \caption{{\footnotesize The motion of the CM in the $XYZ$ frame and in the
          $X'Y'Z'$ frame---a clockwise rotation of $XYZ$ about the $X$-axis by $\theta$.
          The thick straight line represents the trajectory of the CM.}}
 \label{qtounp}
\end{center}
\end{figure}

As we have seen, the path of the CM is a straight line that makes
an angle $\alpha$ with $Y$-axis as shown in figure \ref{qtounp}.
We now introduce a new generalized coordinate: the distance along
this line, $q$. We shall take its zero to correspond to the
position of the CM when the cone is at the bottom of the ramp. In
 figure \ref{qtounp}, $\vec{a}_0$ represents the value of
$\vec{a}$ when the cone is at the bottom of the ramp and $d=0$. It
can be seen that, in terms of $q$, $Y_{CM}$ is given by
\begin{displaymath}
   Y_{CM} \= Y_{CM}^0+q\cos\alpha~,
\end{displaymath}
where $Y_{CM}^0$ is the coordinate of the CM when the cone is at
the bottom of the ramp---that is when $q=0$.  It is also clear
from the figure that
\begin{displaymath}
   Y_{CM}^0 \= a_0\sin\alpha=L\tan\psi\sin\alpha~.
\end{displaymath}

\noindent Combining the results with equations (\ref{aofcp}) and
(\ref{cpofcm}), we obtain
\begin{equation}
  a \= L\tan\psi-q\tan\alpha~.
  \label{aofq}
\end{equation}
The rolling without slipping constraint implies that the  angular
velocity $\omega$ of the cone about the symmetry axis is given by
\begin{displaymath}
  \omega \= -{\dot{q} \over a} \= -{\dot{q} \over L\tan\psi-q\tan\alpha}~.
\end{displaymath}
Thus, we obtain for the kinetic energy
\begin{eqnarray*}
  T&=&{1\over 2}m\dot{q}^2+{1\over 2}I\omega^2
   \={1\over 2}\Big[m + {I \over
  (L\tan\psi-q\tan\alpha)^2}\Big]\dot{q}^2~,
\end{eqnarray*}
where $I$ is the moment of inertia of the cone with respect the
symmetry axis.

 For the potential energy, we are concerned with the vertical height of the CM;
 that is, we must  find the
coordinate $Z_{CM}'$ in terms of $q$, where the primed frame
$X'Y'Z'$ is one that is obtained from the unprimed frame $XYZ$ by
rotating the latter  by an angle $\theta$ about the $X$-axis (see
figure \ref{qtounp}). From the figure we see that
\begin{displaymath}
    Z_{CM}' \= Z_{CM}'^{0}-q\sin(\alpha-\theta)~.
\end{displaymath}
Thus, we have for the potential
\begin{displaymath}
   U \= -mgq\,\sin(\alpha-\theta)~,
\end{displaymath}
where we have set the potential to zero at the initial position of
the cone.

For the cone to roll up the ramp, the potential must be a
decreasing function of $q$, that is
\begin{displaymath}
  \sin(\alpha-\theta) ~>~0 ~.
\end{displaymath}
 With the use of
equation (\ref{alph}) and some trigonometric identities, this
condition may equivalently be written
\begin{equation}
  \tan\theta ~<~ {\tan\phi\tan\psi \over \sqrt{1-\tan^2\phi\tan^2\psi}}~.
\label{upcond}
\end{equation}
Recalling the inequality (\ref{eq:inequality}), we see the
denominator is well-defined and is not singular or imaginary.

We may now write down the energy:
\begin{equation}
  E \= T+U \= {1\over 2}\Big[m+{I\over
  (L\tan\psi-q\tan\alpha)^2}\Big]\dot{q}^2-mgq\sin(\alpha-\theta)~.
\label{energy}
\end{equation}
It should be noted that $q$ may not take on arbitrary values.
Indeed, as we have defined it, $q$ has a minimum value of 0. There
will also be a maximum value at which point the cone falls through
the rails.  This point corresponds to the double cone contacting
the ramp at its very tips.  Hence $a=0$ and, from equation
(\ref{aofq}),
\begin{displaymath}
   0 \leq q \leq  L\,{\tan\psi \over \tan\alpha}~.
\end{displaymath}
In the next section we shall see an unexpected property of the
motion about the fall-through point, $q= L\tan\psi/\tan\alpha$.

%%%%%%%%%%%%%%%%%%%%%%%%%%%%%%%%%%%%%%%%%%%%%%%%%%%%%%%%%%%%%%%%%%%%%%%
\section{A Surprising Feature}

The reader should note the position dependent coefficient in the
inertial term of our expression for energy (\ref{energy}). It is
this term that gives the system unique behavior beyond what has
been discussed so far.

\begin{figure}[th!]
  \begin{center}
  \psfrag{qd}{$\dot{q}$}
  \psfrag{q}{$q$}
  \includegraphics[width=10cm]{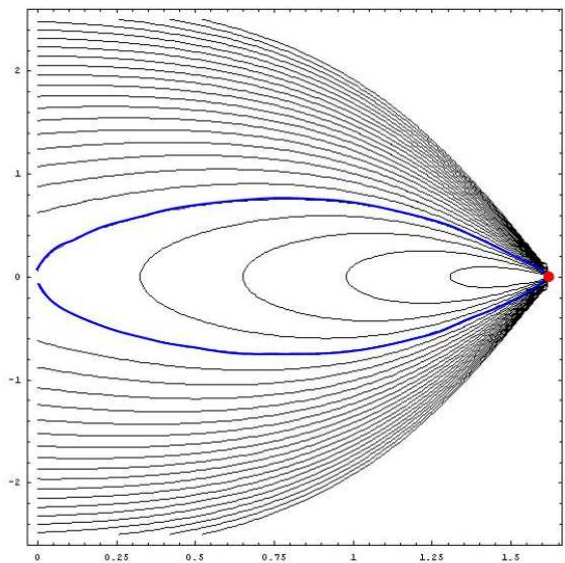}
  \end{center}
  \caption{{\footnotesize $\dot{q}$ vs. $q$ phase portrait for
           $\phi=\psi={\pi \over 6}$, $\theta =
           {\pi\over 12}$, $L=m=1$.}}
  \label{phase}
\end{figure}

Figure \ref{phase} shows the $\dot{q}$ vs. $q$ phase portrait for
the system with $\phi=\psi=\pi/6$, $\theta = \pi/12$, and $L=m=1$.
The first astonishing fact is that \textit{all} of the curves go
through the same point. This is the fall-through point and it will
be the subject a more exhaustive discussion below.

Shown in figure \ref{phase} as a thick curve  is the zero energy
trajectory. Those curves contained within it correspond to
progressively more negative energies. Those exterior,
progressively more positive. One may be tempted to identify, as
cycles, the negative energy trajectories. However, the
fall-through point  is a singularity. While all the trajectories
approach it arbitrarily as $t\to+\infty$, none of them actually
reach it. Thus, the negative energy curves are all open by one
point. Their `open loop' shape merely illustrates the simple fact
that, when the cone is placed at the top of the ramp and given an
initial shove, the cone will roll down for a bit and then roll
back up.

To illustrate just what it is about the motion that is so
startling, let's focus on a single trajectory. Consider the $E=0$
energy curve corresponding to releasing the cone from rest at the
bottom of the ramp (as the demonstration is typically performed).
This predicts that, at first, the cone's translational speed will
be increasing reaching maximum value at about midway along its
path after which the translational speed will begin to ebb,
decreasing to zero at the fall-through point. The cone  will
indefinitely approach but never reach the fall-through point.
Though, at first, hard to accept, with a little thought, one may
see that this behavior does in fact make physical sense.

To make things more concrete let us divide the path of the CM in
to equal tiny increments of, $\Delta q = \epsilon$.  Since the
potential is linear, the cone will gain a fixed amount of energy,
$\delta$, over each increment.  Also, since $a$ is a linear
function of $q$, it will decrease by a fixed amount, $\Delta a$,
over each increment.

Now, let us for the moment, consider a simplified situation; namely,
we shall adjust the inclination of the ramp so that the system is
potentially neutral and have a physicist take over gravity's job. We
shall stipulate:
\begin{enumerate}
 \item  the cone will be given an initial speed $v$;
 \item the physicist will provide and only enough energy over each increment so
       as to keep the cone translating as speed $v$;
 \item the physicist shall not give the cone more than $\delta$ energy per increment.
\end{enumerate}
From the rolling without slipping constraint we may write the
cone's rotational energy as $T_{rot}={1\over 2}I{v^2\over a^2}$.
Differentiating we obtain
\begin{equation}
 \Delta T_{rot} = {1\over 2}\, I\,{v^2\over a^3} \, \Delta a~,
 \label{enin}
\end{equation}
for the amount of energy needed over each increment to keep the
cone translating at speed $v$. Since $a$ goes from a maximum to
zero, we can always find a position along the ramp for which this
amount exceeds $\delta$.  At this point, the physicist will fail
to maintain the cone at its current translational speed. However,
the physicist may transfer some of the translational energy into
rotation, causing the cone to slow down and repeat the process for
each of the following intervals.

Returning to the actual situation, with the cone rolling freely on
the ramp, it is the same principle that is responsible for the
cone's peculiar motion. However, the accountant that keeps track
of energy transactions is no longer the physicist but gravity.
Also, in this case, energy is initially being proportioned between
the translational and rotational energy.  We may for the sake of
argument, use the speed of the cone, say, one microsecond after
release, as the $v$ in equation (\ref{enin}).  No matter how small
it is, we may still find a position for which the right hand side
of equation (\ref{enin}) exceeds $\delta$. In reality the initial
increase of the speed of the cone as it moves along only serves to
bring about the cross-over point sooner. As seen in figure
\ref{phase}, the higher the energy, the more the curve `flattens'
at the beginning, eventually loosing the increasing part
altogether.

By straightforward extension of the above argument, we thus, see
more and more energy has to be transferred form translation to
rotation as $a$ shrinks down to zero.  This results in an
asymptotic behavior of pure rotation about the symmetry axis. The
energy transfer mechanism in all of this is ultimately
friction---which we assume to be ideal---working to keep the cone
from slipping.

%%%%%%%%%%%%%%%%%%%%%%%%%%%%%%%%%%%%%%%%%%%%%%%%%%%%%%%%%%%%%%%%%%%%%%%%%%%%%
\section{Integration of the Equation of Motion}

 Equation (\ref{energy}) can be solved for the velocity,
\begin{equation}
  \dot{q} \= \pm\sqrt{{2(E+mgq\sin(\alpha-\theta))\over m+{I\over
  (L\tan\psi-q\tan\alpha)^2}}}~.
  \label{qdot}
\end{equation}
Notice that, except for in the degenerate cases in which $\phi$ or
$\psi = 0$ (and thus also $\alpha=0$),
\begin{displaymath}
  \lim_{ q \rightarrow L{tan\psi \over \tan\alpha}}\dot{q} \= 0~.
\end{displaymath}
Thus the ebb in the cones speed is a general feature.

\begin{figure}[h!]
\begin{center}
 \includegraphics[width=7cm]{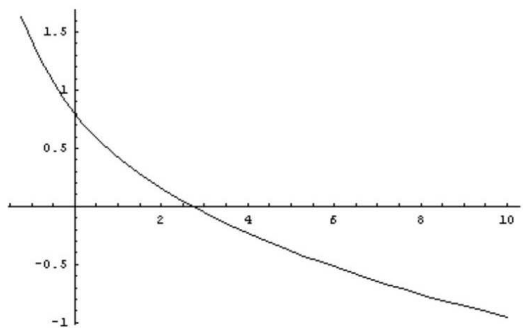}
 \caption{Cross-over point vs. Energy}
 \label{turning}
\end{center}
\end{figure}

Equation (\ref{qdot}) gives the velocity of the CM as a function
of position, $\dot q=\dot q(q)$. We can find where the cross-over
point occurs by taking the derivative of this equation and setting
it equal to zero.  The resulting expression is quite turbid. More
meaningful is a plot of the cross-over point versus energy. This
is shown in figure \ref{turning} where we used the same values of
the parameters used for the  phase plot of figure \ref{phase}.

Notice that the cross-over point is a decreasing function of
energy. As we have already pointed out, increasing energy means
increasing the starting velocity which brings the cross-over point
closer. Also interesting is that, above a certain energy, the
cross-over point occurs at negative values of $q$. For large
enough energy, the initial velocity is so high that gravity can
never provide enough energy over an increment to keep the cone
translating at its current speed. Therefore at these energies the
speed will be a strictly decreasing function of time. This may be
quickly confirmed using the extreme case $E\to+\infty$ in
(\ref{qdot}).

We may now integrate equation (\ref{qdot}) to get
\begin{equation}
  \beta\pm t \= \int  \sqrt{{m+{I\over
   (L\tan\psi-q\tan\alpha)^2}\over 2(E +
    mgq\sin(\alpha-\theta))}}\, dq~,
\label{eom}
\end{equation}
where $\beta$ is a constant of integration. There is no simple
analytical solution to this integral. We may however make
approximations in two regimes: small values of $q$ (cone close to
the vertex) and values of $q$ close to $L\tan\psi/\tan\alpha$
(cone close to the fall-through point).

%%%%%%%%%%%%%%%%%%%%%%%%%%%%%%%%%%
\subsubsection{Near Vertex Regime}

\begin{figure}
\begin{center}
 \includegraphics[width=6cm]{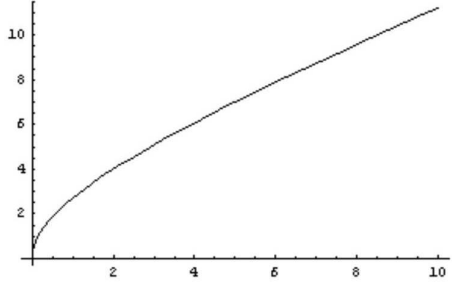}
 \caption{The graph of $t$ vs $q$ close to the vertex of the ramp.}
 \label{sme0}
\end{center}
\end{figure}

For small values of $q$, we may apply the binomial expansion to the
${I\over (L\tan\psi-q\tan\alpha)^2}$ term in the numerator of
(\ref{qdot}) and approximate the integral as
\begin{displaymath}
  \beta\pm t\approx\int\sqrt{{A+Bq\over E+Cq}}\,{dq\over2}~,
\end{displaymath}
where
\begin{eqnarray*}
  A ~\equiv~ m+{I\over L^2\tan^2\psi}~,~~~
  B ~\equiv~ {2I\tan\alpha \over L^3 \tan^3\psi}~,~~~
  C ~\equiv~ mg\sin(\alpha-\theta)~.
\end{eqnarray*}
The last integral is elementary and gives
\begin{eqnarray*}
  \beta\pm t&\approx&{\sqrt{{A+Bq\over E+Cq}}(E+Cq)\over C}\\
            &&+{(AC-BE)\sqrt{{A+Bq\over E+Cq}}\sqrt{E+Cq}
            \ln\big(2\sqrt{(A+Bq)(E+Cq)}+{BE+AC+2BCq\over
            \sqrt{BC}}\big)\over 2C^{3/2}\sqrt{B(A+Bq)}}~.
\end{eqnarray*}
Figure \ref{sme0} plots the solution for the example that we have
been using throughout, with initial conditions $q=0$ and
$\dot{q}=0$ at $t=0$. The inverse curve $q=q(t)$, is found by
reflection with respect to the diagonal $q=t$ of the unit
`square'. Notice that the figure indicates that the cone is
speeding up.

If one makes an additional binomial expansion on equation
(\ref{eom}), a simpler expression is obtained
\begin{displaymath}
q(t)\approx{e^{\sqrt{E(\beta\pm
t)}}\big(B^2-4AE-2B\sqrt{E}e^{\sqrt{E(\beta\pm
t)}}+Ee^{2\sqrt{E(\beta\pm t)}}\over 4E^{3/2}}~.
\end{displaymath}

%%%%%%%%%%%%%%%%%%%%%%%%%%%%%%%%%%%%%%%
\subsubsection{Near Fall-Through Regime}
The integral of equation (\ref{eom}) may be written in terms of
$a$ as
\begin{displaymath}
   -\beta\pm t = \int\cot\alpha\, \sqrt{{m+{I\over a^2}\over D -
       Fa}}\,da~,
\end{displaymath}
where
\begin{eqnarray*}
     D ~\equiv~
     2E+2mgL\, {\sin(\alpha-\theta)\,\tan\psi\over\tan\alpha}~,~~~
     F ~\equiv~ 2mg\sin(\alpha-\theta)\cot\alpha~.
\end{eqnarray*}
For, $q\approx L{\tan\psi\over\tan\alpha}\Rightarrow a\approx
0$,
\begin{displaymath}
  -\beta\pm t \approx \sqrt{{I\over D}}\,\cot\alpha\,\int{da \over a}~.
\end{displaymath}
Integrating and inverting, we obtain
\begin{displaymath}
 q(t)\approx L{\tan\psi\over\tan\alpha}-\cot\alpha \,
 \exp\Bigg[\tan\alpha\sqrt{{D\over I}}(-\beta\pm t)\Bigg]~.
\end{displaymath}

\begin{figure}[hb!]
\begin{center}
\includegraphics[width=6cm]{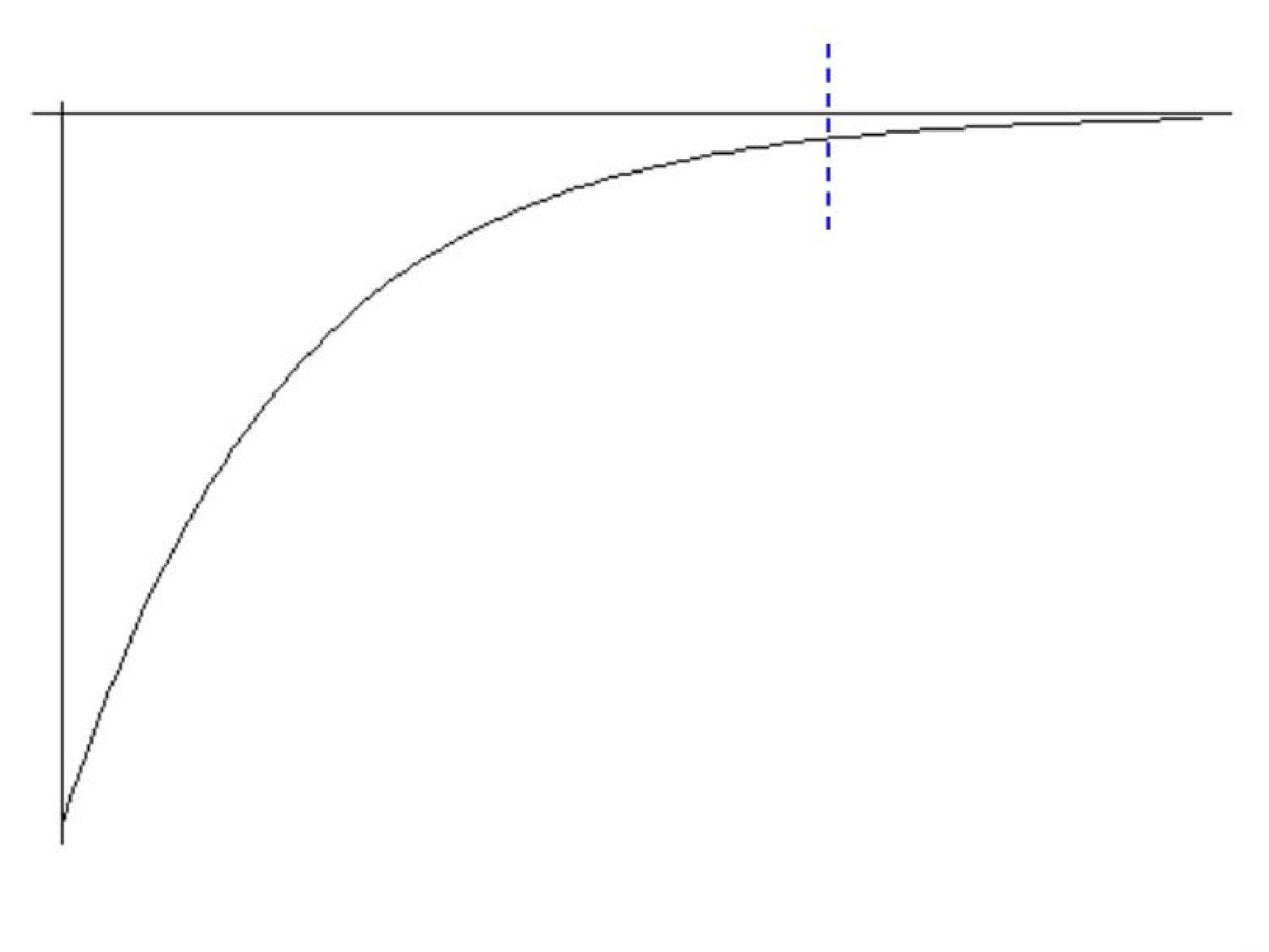}
\hspace{1cm}
 \includegraphics[width=5.7cm]{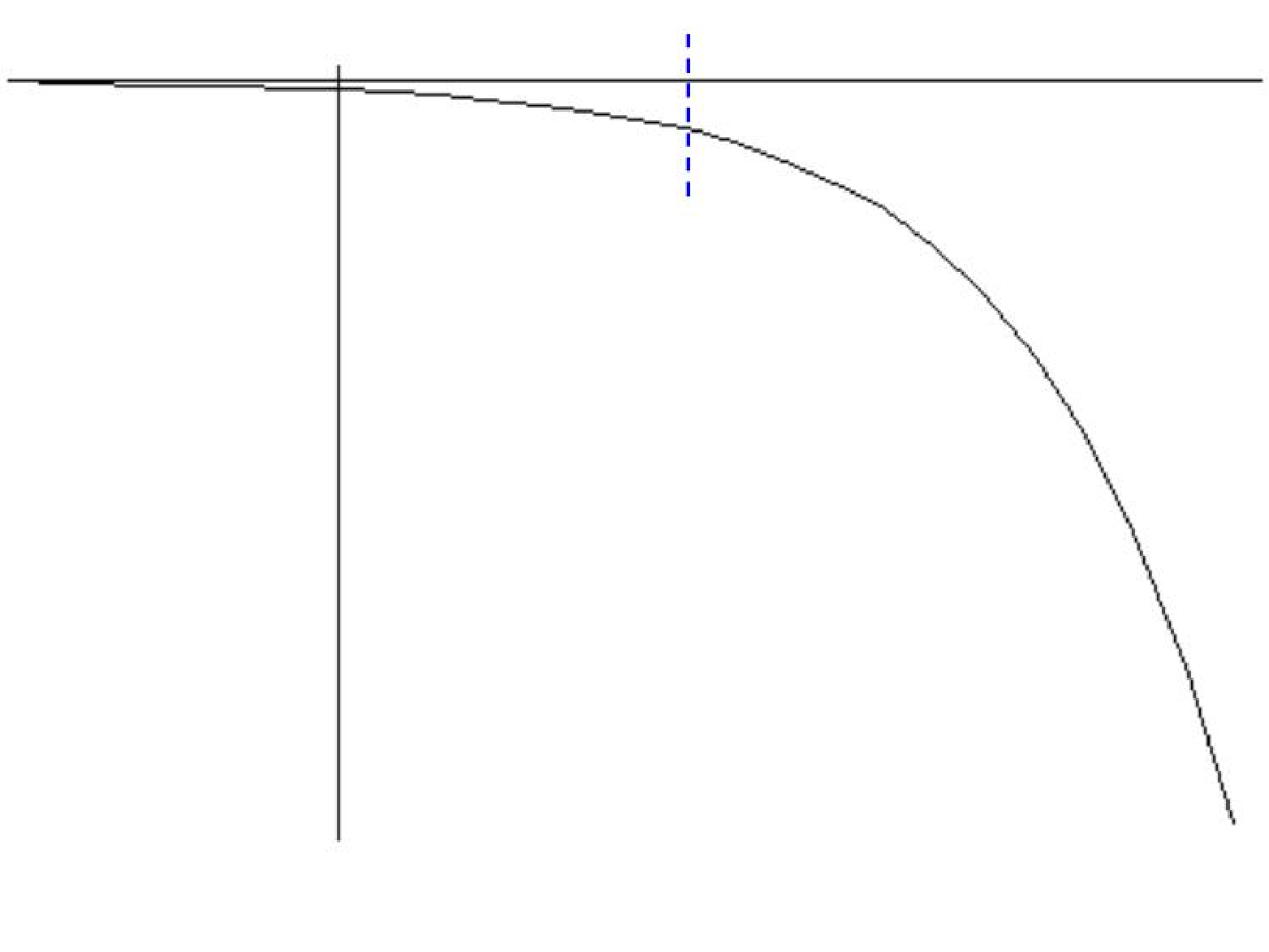}
 \caption{{\footnotesize $q$ vs $t$ near the fall-trough point.}}
 \label{negt}
\end{center}
\end{figure}

The general form of the solution is that of the left side of
figure \ref{negt}, when the minus sign is chosen in the above
expression, representing the cone moving up the ramp toward the
fall through point.  It is that of the right side figure
\ref{negt}, when the positive sign is chosen, representing the
cone moving down the ramp, starting at some point very close to
the fall through point at $t=0$.  Notice that the graph on the
left depicts the cone speeding up as it gains potential and that
on the right slowing down as it drops in potential. In the
figures, the dotted vertical lines roughly mark off the regions in
which the solutions are valid. The vertical axes are at
$q=L{\tan\psi\over \tan\alpha}$, the point at which the cone falls
through the rails. Due to this  asymptotic behavior, it would take
infinite time for the cone to reach the fall-through point or
reach any other point starting from the fall-through point.

\begin{figure}[h!]
\begin{center}
 \psfrag{s}{\textcolor{blue}{$\psi$}}
 \includegraphics[width=7cm]{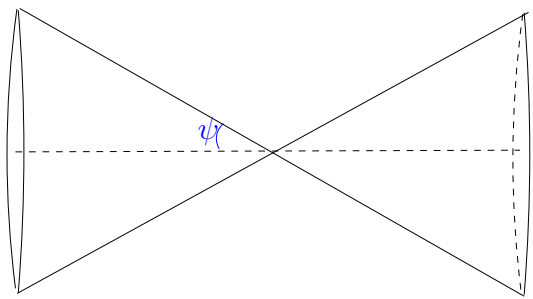}
 \caption{{\footnotesize The simplest variation of the double-cone demo
           would be an hourglass
          sitting on a V-shaped ramp that has its vertex elevated.}}
\label{hour}
\end{center}
\end{figure}

%%%%%%%%%%%%%%%%%%%%%%%%%%%%%%%%%%%%%%%%%%%%%%%%%%%%%%%%%%%%%%%%%%%%%%%%%%
\section{The Hourglass}
It is not difficult to invent variations on this system (see, for
example \cite{TPT}). In particular, one may obtain an essentially
identical motion by `inverting' the double-cone as well as our
ramp---that is, flipping the ramp up side down so that its vertex
is at the top, and placing upon it an hourglass-shaped object such
as the one depicted in figure \ref{hour}.  If we define our
parameters in a similar manner as before (adjusting, as in the
figure, the definition of $\psi$ appropriately), and redefine $q$
so that its zero now corresponds to the location of the CM when
the hourglass is at the top of the ramp and so that it increases
in the direction in which the cone travels down the ramp, we
obtain for the energy,
\begin{displaymath}
  E \={1\over 2}\Big[m+{I\over
  q^2\sec^4\alpha\tan^2\alpha}\Big]\dot{q}^2+m g q\sin(\alpha-\theta)~.
\end{displaymath}
The result can be seen to posses essentially the same form as the
energy of the double cone, and thus the system's motion will be
similar, with the cone, now, tending to roll in the direction of
decreasing $q$--again, up the ramp--and the singularity, now,
occurring at $q=0$.

\section{Concluding Remarks}
We see that the implications of the unique geometry of this system
extend far beyond the simple conceptual explanation relying on a
rough trend of the CM.  Despite the simplicity of the demo, the
dynamical explanation, in terms of the consequences of the cone's
peculiar way of sitting on the ramp, has revealed surprising
properties. We hope that instructors of introductory physics will be
as charmed as we are by this demo and its rich physical content.

\section*{Acknowledgements}
This research was sponsored by generous grants from the University
of Central Florida Honors College and Office of Undergraduate
Studies. S.G. would like to thank Rick Schell and Alvin Wang for
this financial support.  S.G. would like to thank C.E. for the
opportunity to work on this project, his generous time and
guidance throughout this work and S.G.'s studies.

%%%%%%%%%%%%%%%%%%%%%%%%%%%%%%%%%%%%%%%%%%%%%%%%%%%%%%%%%%%%%%%%%%%%%%

\end{document}